\documentstyle[prb,aps, amssymb, amsfonts, amsmath, 12pt]{revtex}

\tighten

\renewcommand{\thanks}[1]{\renewcommand{\thefootnote}{\rm\alph{footnote})}%
   \protect\footnote{#1}}

\makeatletter
\title{Lattice Approximation of Quantum Statistical
Traces at a Complex Temperature}
\author{
J.~Lukkarinen\thanks{E-mail address: jani.lukkarinen@helsinki.fi}}
\address{ Helsinki Institute of Physics, P.O.Box 9, 
00014 University of Helsinki, Finland} 
\makeatother

\newcommand{\braket}[2]{\langle #1|#2\rangle}

\newcommand{\vc}[1]{{\bf #1}}

\newcommand{\tr}{{\rm Tr}\,}

\newcommand{\re}{{\rm Re\,}}

\newcommand{\ope}[1]{\widehat{#1}}
\newcommand{\norm}[1]{\Vert #1\Vert}
\newcommand{\esssup}{ {\rm ess\,sup} }

\newcommand{\ham}{\ope{H}}
\newcommand{\hp}{\ope{H}_0}
\newcommand{\Rn}{{\mathbb R}^n}
\newcommand{\TN}{\ope{T}({t/ N})^N}
\newcommand{\T}{\ope{T}}
\newcommand{\iker}[4]{#1(\vc{#2},\vc{#3}; #4)}
\newcommand{\Ptker}[2]{\iker{P}{#1}{#2}{t}}
\newcommand{\knker}[2]{\iker{K_N}{#1}{#2}{t}}
\newcommand{\kiker}[2]{\iker{K_\infty}{#1}{#2}{t}}

\newtheorem{theorem}{Theorem}[section]

\newtheorem{corollary}[theorem]{Corollary}
\newtheorem{lemma}[theorem]{Lemma}
\newenvironment{proof}{\begin{trivlist}\item[]{\em 
Proof:}\/}{\hfill\mbox{$\Box$}\end{trivlist}}

\begin{document} 

\preprint{HIP-1997-58/TH}
 
\maketitle

\begin{abstract}
We prove that the simple condition on the potential $V$,
$\int \exp(-t V) < \infty$ for all $t>0$, 
is sufficient for the lattice approximation of
$\tr(\ope{A}e^{-\beta\ham})$ with $\re\beta >0$ to work for all bounded
functions $A$ and a large class of potentials.  As a by-product
we obtain an explicit bound for the real-temperature lattice kernels.
\end{abstract}


\section{Introduction}
Ever since the introduction of the path-integrals by R.~Feynman, the
path-integral formulas with a complex ``temperature'' have presented
an elusive mathematical problem---for instance, 
the Wiener measure, which is so
useful for positive temperature statistics, cannot be extended to the complex
case.\cite{cam62} 
In fact, since real time processes and real temperature statistics
can be identified with each other via analytic continuation, 
the present numerical non-perturbative 
methods in quantum field theories rely mostly on
computations with lattice approximations to
the {\em imaginary time}\/,
i.e.\ real temperature, path-space measures.

There have already been proposals for numerical
evaluation of the complex path-integrals\cite{klau84,par83}, but
only now can we establish rigorously the assumed relations
between the complex lattice integrals and the continuum traces.  Also,
the result of Theorem \ref{th:complexb} has been used\cite{jl:97} as
an assumption in the derivation of a lattice approximation 
to the quantum
microcanonical ensemble and the theorem now yields a wide range of
potentials for which this approximation is valid.

For an introduction to the mathematics of operators on a Hilbert space
see e.g.\ the second part of the comprehensive series\cite{reedsimonII}
by Reed and Simon.  The path-space measures and
their applications are discussed in detail in 
another book\cite{simon:fint} by Simon and the 
proofs of the statements made in the following section can be found from
Section 6 of that book.

\section{Notations and Definitions\label{sec:defnot}}

We will now consider quantum mechanics 
on the separable Hilbert space $L^2(\Rn)$; in the following, $n$ will
always denote the number of dimensions of the parameter space and
we will use boldface letters for vectors of $\Rn$.
For simplicity, we have also adopted the following non-standard
notations: $D\vc{x}$ denotes the n-dimensional Lebesque measure 
with respect to the variable $\vc{x}$ and $D^N\vc{x}$ stands for the
product measure $\prod_{k=1}^N D\vc{x}_k$.  Operators will be
distinguished by the ``hat''---especially the multiplication operator
defined by a measurable function $F$ will be denoted by
$\ope{F}$. 

The free Hamiltonian is defined in the standard way,
\[
\hp := { \ope{p}^2 \over 2 m} = -{\hbar^2\over 2 m}\nabla^2,
\]
with the usual domain for this 
operator.  To simplify the following discussion we will from now on use 
the natural units $\hbar=1$ and, in addition, consider only the case
$m=1$.  The results for the situation $m\ne 1$ are easily reproduced
from the following formulas by simply replacing $t$ by $t/m$ and the
potential $V$ by $m V$.  

It is well-known that $\hp$
generates an analytic semigroup on the right half-plane, 
where $e^{-\beta\hp}$ are bounded
integral operators for $\re\beta >0$ and they are 
given by the integral kernels
\[
\iker{P}{a}{b}{\beta} :=  {1
  \over \left( 2 \pi \beta \right)^{n/2} }
  \exp(-{1\over 2 \beta} |\vc{a}-\vc{b}|^2).
\]
It is also evident that these integral kernels satisfy the semigroup
property exactly, i.e.\ that for all $\vc{a}$, $\vc{b}\in\Rn$ and
$\re \beta_1$, $\re \beta_2>0$,
\[
\int\!\! D\vc{c}\, \iker{P}{a}{c}{\beta_1} 
  \iker{P}{c}{b}{\beta_2} = \iker{P}{a}{b}{\beta_1+\beta_2}.
\]

The potential $V$ is assumed to be locally in $L^1(\Rn)$ and positive, 
but the results easily generalize for all real $L^1_{\text{loc}}$
potentials bounded from below (just replace $V$ by
$V+\esssup(-V)$). The Hamiltonian $\ham=\hp+\ope{V}$ 
can now be defined as a sum of quadratic 
forms, when it is self-adjoint and bounded
from below and, therefore, $e^{-\beta\ham}$ 
(as well as $e^{-\beta\ope{V}}$)
is an analytic semigroup on the right half-plane.  In addition, we shall
also require that
\begin{equation}\label{e:Vassum}
\int\!\! D\vc{a}\, e^{-t V(\vc{a})}< \infty\mbox{, for all }t>0,
\end{equation}
so that $e^{-t\ham}$ is trace-class for $t>0$ by the
Golden, Thompson, Symanzik -inequality\cite{gtz}.

Under these assumptions $e^{-t\ham}$ is an integral
operator for $t>0$ and it has an integral kernel $K_\infty$,
which can be obtained with aid of
the conditional Wiener measure $\mu$: for
all $\vc{a}$, $\vc{b}\in\Rn$ and for all $t>0$ define
\[
\kiker{a}{b} = \int\!\!d\mu_{\vc{a}, \vc{b}; t}(\omega)\, e^{-\int_0^t ds
 V(\omega(s))}.
\]
In addition to (\ref{e:Vassum}) we shall now require the potential 
$V$ to be such that the Wiener integrals can almost everywhere
be obtained from the lattice kernels
\[
\knker{a}{b} :=  \int\! D^{N-1}\vc{x} 
  \left( {N\over 2 \pi t} \right)^{\textstyle {N n\over 2} } \!\!
  \exp\!\!\bigg[ -{N\over 2 t}\sum_{k=1}^N |\vc{x}_{k-1}-\vc{x}_k|^2
  -{t\over N} \sum_{k=1}^N V(\vc{x}_k)
 \bigg]_{\overset{\scriptstyle\vc{x}_0=\vc{a}\hfill}{\vc{x}_N=\vc{b}}},
\]
i.e.\ that
\begin{equation}\label{e:latlim}
 \kiker{a}{b} = \lim_{N\to\infty} \knker{a}{b}\mbox{, for all }t>0
  \mbox{ and for almost all }\vc{a}, \vc{b}\in\Rn.
\end{equation}
This requirement is clearly satisfied for {\em all}\/ 
$\vc{a}$ and $\vc{b}$ if $V$ is continuous, since then
$\int\! ds\, V(\omega(s))$ can be approximated by suitable Riemann sums
leading to (\ref{e:latlim}) by 
using the definition of the Wiener measure.

We shall also need the operator
\[
\ope{T}(\beta) := e^{-\beta\hp} e^{-\beta\ope{V}},
\]
which is clearly a bounded 
integral operator for $\re\beta > 0$, since then $\norm{\T(\beta)} \le
\norm{e^{-\beta\hp}} \norm{e^{-\beta\ope{V}}} \le 1$.  Notice that 
$\iker{K_N}{\cdot}{\cdot}{\beta}$ 
is nothing but the integral kernel of the operator
$\ope{T}\big({\beta\over N}\big)^N$.

\section{Two Lattice Theorems}

In this section we will prove two theorems concerning the convergence of
traces of lattice operators.  The first theorem will show that
also certain unbounded multiplication operators are measurable using
the (real-temperature) Gibbs statistics, if the
potential increases sufficiently fast at infinity.  This will be proven
by using the explicit bounds given by the following lemma and its two
corollaries.  The rather technical proofs of the lemma and the corollaries are
included in Section \ref{SEC:LEMPROOFS}---the lemma itself is an
extension of a result derived by Symanzik\cite{sym65} for the continuum
kernels. 
\begin{lemma}\label{lem:knl}
For all $\vc{a}$, $\vc{b}$ in $\Rn$, for all $t>0$ and for all positive
integers $N$,
\begin{equation}\label{e:knl}
\begin{split}
\knker{a}{b} \le
 &\ C_n t^{n-1\over 2}\!
 \int\!\! D\vc{c}\, \Ptker{a}{c} e^{-t V(\vc{c})} \Ptker{c}{b}
 \left( {1\over |\vc{a}-\vc{c}|^{n-1}} + {1\over |\vc{c}-\vc{b}|^{n-1}}
  \right) \\
  & \quad + {1\over N} e^{-t V(\vc{b})} \Ptker{a}{b},
\end{split}
\end{equation}
where $C_n$ depends only on the dimension $n$.
\end{lemma}

Taking the limit $N\to\infty$ in the previous lemma shows that
{\em
(\ref{e:knl}) is valid for almost all $\vc{a}$, $\vc{b}$ 
and for all $t>0$ also when $N=\infty$.}

\begin{corollary}\label{cor:intbounds}
Assume that in addition to (\ref{e:Vassum})
the potential satisfies
\[
\int\!\! D\vc{a}\, |\vc{a}|^r e^{-t V(\vc{a})} < \infty,
\]
for some $r\ge 0$ and for all $t>0$.  Then
there exists functions $F_r$ and $G_r$, continuous on $(0,\infty)$,
such that for all $t>0$ and for all $N<\infty$,
\[
\int\!\! D\vc{a}\, |\vc{a}|^r \knker{a}{a} \le F_r(t)
\]
and for all $N$, including $\infty$,
\[
\int\!\! D\vc{a}D\vc{b}\, |\vc{a}|^r \knker{a}{b} \le G_r(t).
\]
\end{corollary}

If the above assumption on the potential is true for 0 and $r$
then it is clearly true for all values $s$ in $(0,r)$ 
by the inequality $|\vc{a}|^s \le 1 + |\vc{a}|^r$.  If the potential
has no singularity in a neighborhood of the origin, it is enough
to find one $r\ge 0$ for the statement to be true for all $s$ with
$0\le s\le r$.

\begin{corollary}\label{cor:supbound}
For all $N$, including $\infty$, and for all $t>0$, 
$\underset{\vc{a},\vc{b}\in\Rn}{\esssup}\,
 \knker{a}{b} \le C'_n t^{-{n\over2}}$.
\end{corollary}

The convergence properties of the lattice traces are easily derived from
these results; we have only to be careful that the limit of the traces
of the integral kernels does give the trace of the continuum
integral operator. 
This leads to the appearance of the factor of two in Eq.\
(\ref{e:tr2k}), which will be discussed more after the proof of the
theorem. 

\begin{theorem}\label{th:realt}
Let $A$ be measurable
function bounded by a degree $r\ge 0$ polynomial and assume that the
potential additionally satisfies
\[
\int\!\! D\vc{a}\, |\vc{a}|^{2 r} e^{-t V(\vc{a})} < \infty, 
\text{ for all }t>0.
\]
Then for all $t>0$ and $N\ge 2$ the operators
$\ope{A} e^{-t\ham}$ and $\ope{A} \TN$
are both Hilbert-Schmidt and trace-class and
\begin{gather}
\tr\!\!\left(\ope{A} e^{-t\ham}\right) 
   = \lim_{k\to\infty} \tr\!\!\left(\ope{A} 
                 \T\Big(\smash[t]{t\over 2 k}\Big)^{2 k}\right), 
         \label{e:tr2k}\\
\tr\!\!\left(\ope{A} \TN\right)
   = \int\!\! D\vc{a}\, A(\vc{a}) \knker{a}{a}.
\end{gather}
\end{theorem}
\begin{proof}
The requirement for $A$ is equivalent to finding a
$C\ge 0$ such that $|A(\vc{a})|\le C (1+|\vc{a}|^r)$ for all $\vc{a}$.
It follows now from Corollaries \ref{cor:intbounds} and
\ref{cor:supbound} that for $1\le N \le \infty$,
\[
\int\!\! D\vc{a}D\vc{b}\, |A(\vc{a}) \knker{a}{b}|^2 < \infty.
\]
The operators $\ope{A} e^{-t\ham}$ and $\ope{A} \TN$
are given by the integral kernels $A(\vc{a}) \knker{a}{b}$ 
and, therefore, they are Hilbert-Schmidt.

Since $\ope{A} e^{-t\ham}$
is a product of the two Hilbert-Schmidt operators 
$\ope{A} e^{-{t\over 2}\ham}$ and $e^{-{t\over 2}\ham}$, 
it is trace-class.  Moreover, the trace can be expressed
in terms of the integral kernels of these Hilbert-Schmidt operators:
\begin{align*}
\tr\!\!\left(\ope{A} e^{-t\ham}\right) = 
  \int\!\! D\vc{a}D\vc{b}\, A(\vc{a}) \iker{K_\infty}{a}{b}{{t\over2}}
      \iker{K_\infty}{b}{a}{{t\over2}}.
\end{align*}
Using Lemma \ref{lem:knl} and Corollary \ref{cor:supbound}
to give an upper bound for the Lebesque
dominated convergence theorem shows now that
\begin{gather*}
\begin{split}
\int\!\! D\vc{a}D\vc{b}\, A(\vc{a}) \iker{K_\infty}{a}{b}{{t\over2}}
    \iker{K_\infty}{b}{a}{{t\over2}} &=
\lim_{N\to\infty} \int\!\! D\vc{a}D\vc{b}\, A(\vc{a})
 \iker{K_N}{a}{b}{{t\over2}} \iker{K_N}{b}{a}{{t\over2}} \\
 & = \lim_{N\to\infty} \int\!\! D\vc{a} \, A(\vc{a})\iker{K_{2 N}}{a}{a}{t},
\end{split}
\end{gather*}
where the second equality follows from Fubini's theorem and a careful
inspection of the definition of $K_{2 N}$.  Therefore,
\begin{equation}\label{e:trdres}
\tr\!\!\left(\ope{A} e^{-t\ham}\right) =
\lim_{N\to\infty} \int\!\! D\vc{a}\, A(\vc{a}) \iker{K_{2 N}}{a}{a}{t}.
\end{equation}

It has already been proven that both $\ope{A} \ope{T}(t)$ 
and $\ope{T}(t)^{N-1}$ are 
Hilbert-Schmidt for all $t>0$ and thus
$\ope{A} \TN$ is
trace-class and its trace can be expressed in terms of the kernels of
these operators.  It is then a matter of inspection to conclude
\begin{equation}\label{e:trndres}
\tr\!\!\left(\ope{A} \TN\right) = 
  \int\!\! D\vc{a}\, A(\vc{a}) \knker{a}{a}
\end{equation}
and the theorem follows from equations (\ref{e:trdres}) and
(\ref{e:trndres}). 
\end{proof}

If $\lim_N \knker{a}{a}$ exists for almost all $\vc{a}$, then an
application of the dominated convergence theorem in (\ref{e:trdres})
shows that $\tr\!\!\left(\ope{A} e^{-t\ham}\right) =
 \lim_N \int\!\! D\vc{a}\, A(\vc{a}) \knker{a}{a}$.  Therefore, 
also odd lattice sizes can be included in the limit, for instance,
if $V$ is continuous (then $K_\infty$ is continuous and thus
finite on the diagonal).

In the second theorem we will work with temperatures having an imaginary
part.  The following proof relies on a simple property of the
Hilbert-Schmidt operators given as Lemma \ref{lem:trh} in Section 
\ref{SEC:LEMPROOFS}.  To make the theorem easier to use,
we will now state all the assumptions made on the potential explicitly.

\begin{theorem}\label{th:complexb}
Assume that $A$ is an essentially bounded measurable function and 
$V\in L^1_{\text{loc}}$ is a potential bounded from below for which
(\ref{e:latlim}) holds and
\[
\int\!\! D\vc{a}\, e^{-t V(\vc{a})} <\infty \mbox{, for all }t>0.
\]
Then for all complex $\beta$ with $\re\beta > 0$, 
$\ope{A} e^{-\beta\ham}$ is trace-class and 
\begin{gather*}
\tr\!\!\left(\ope{A} e^{-\beta\ham}\right)
   = \lim_{N\to\infty}\int\!\! D\vc{a}\, A(\vc{a}) 
      \iker{K_{2 N}}{a}{a}{\beta}.
\end{gather*}
\end{theorem}
\begin{proof}
Let $f_N(\beta)=\int\!\! D\vc{a}\, A(\vc{a}) \iker{K_{2 N}}{a}{a}{\beta}$
and $g(\beta)=\tr\!\!\left(\ope{A} e^{-\beta\ham}\right)$.  Both are
then well-defined
on the right half-plane: $f_N$ is an $2 N$-fold integral
over an absolutely integrable function and, 
by the Schwarz inequality,
self-adjointness of $\ham$ and Theorem \ref{th:realt},
\[
|g(\beta)|^2 \le \tr\!\!\left(|\ope{A}|^2 e^{-\re\beta\ham}\right)
  \tr\!\!\left(e^{-\re\beta\ham}\right) < \infty.
\]
In addition, $g$ and all $f_N$ are analytic on the right
half-plane, as can be seen e.g.\ by a 
standard application of Morera's theorem---note that 
\[
g(\beta) = \sum_n \braket{\psi_n}{\ope{A}\psi_n} e^{-\beta E_n},
\]
where $\{E_n\}$ are the eigenvalues of $\ham$ and $\{\psi_n\}$ is a
corresponding orthonormal basis.

Since Theorem \ref{th:realt} states that 
$\lim_{N\to\infty} f_N(\beta)=g(\beta)$ for positive
$\beta$, the theorem will follow from the Vitali convergence 
theorem\cite{vitalict} if we can only
prove that the sequence $(f_N)$ is uniformly bounded on every compact
subset of the half-plane.

An argument similar to one in the proof of Theorem \ref{th:realt} shows that
$f_N(\beta)=\tr\!\!\left(\ope{A} 
\T({\beta\over 2 N})^{2 N}\right)$.  Therefore, Lemma
\ref{lem:trh} establishes the bound
$|f_N(\beta)| \le \norm{\ope{A}} \tr(\T^*\T)^N$, where
$\T=\T({\beta\over 2 N})$.  However, as $\T^*\T = 
e^{-{\beta^*\over 2 N}\ope{V}} 
e^{-{\re\beta\over N} \hp} e^{-{\beta\over 2 N}\ope{V}}$,
\[
\tr(\T^*\T)^N = \tr\!\!\left(\T(\re\beta/N)^N\right) \le 
  F_0(\re\beta) < \infty
\]
by Corollary \ref{cor:intbounds} and Theorem \ref{th:realt}.
If $S$ is a compact set in the right half-plane, it follows that
$F_0(\re S)$ is
bounded since $F_0$ was continuous.  This proves that $(f_N)$ is uniformly
bounded on every compact subset of the right half-plane and hence
completes the proof.
\end{proof}

If $V$ is continuous, then the limit can be taken by using all values of
$N$:  Define $h_N(\beta)=\int\!\! D\vc{a}\, A(\vc{a})
\iker{K_N}{a}{a}{\beta}$.  By the remark after Theorem \ref{th:realt},
$\lim_N h_N = g$ on the positive real axis.  Also,
\[
 |h_{2 N+1}(\beta)| = \left|\tr\!\!\left(\ope{A}
\T({\beta\over 2 N +1 })^{2 N+1}\right)\right| \le
 \norm{\ope{A}} \norm{\T} \tr(\T^*\T)^N \le \norm{\ope{A}} F_0(\re\beta
      {2 N \over 2 N +1}),
\]
and thus the sequence $(h_N)$ is uniformly bounded on the right half-plane.

\section{Proofs of the Lemmas\protect\label{sec:lemproofs}}

{\em Proof of Lemma \ref{lem:knl}:}\/
By the relation between geometric and arithmetic means,
$\exp(-{t\over N} \sum_{k=1}^N V(\vc{x}_k)) \le 
   {1\over N} \sum_{k=1}^N \exp(-t V(\vc{x}_k))$.  Using this and
the semigroup property of the integral kernels $P$, 
we arrive at the inequality
\[
\knker{a}{b} \le {1\over N} e^{-t V(\vc{b})} \Ptker{a}{b}
 + \int\! D\vc{c}\, e^{-t V(\vc{c})} 
    {1\over N} \sum_{k=1}^{N-1} \iker{P}{a}{c}{ k {t\over N} } 
    \iker{P}{c}{b}{ (N-k){ t\over N} }.
\]
The terms in the last sum are explicitly
\begin{align}
\begin{split}
\bigg({1\over 2 \pi t}& {N \over \sqrt{k (N-k)} }\bigg)^{\textstyle n}
 \exp\!\!\bigg[-{1\over 2 t} \bigg( {N\over k} |\vc{a}-\vc{c}|^2 + {N\over N-k}
    |\vc{c}-\vc{b}|^2 \bigg) \bigg]
\end{split} \notag \\
\begin{split}
\qquad &= \Ptker{a}{c} \Ptker{c}{b} {N \over \sqrt{k (N-k)} }
  \bigg( {N-k\over k} \bigg)^{\textstyle {n-1\over 2} }
  \exp\!\!\bigg[-{N-k\over k}
   { |\vc{a}-\vc{c}|^2 \over 2 t} \bigg] \\
&\qquad \times
  \bigg( {N\over N-k} \bigg)^{\mbox{$n$$-$$1$} }
    \exp\!\!\bigg[ - {k\over N-k} 
       { |\vc{c}-\vc{b}|^2 \over 2 t} \bigg]
\end{split} \label{e:1steq} \\
\begin{split}
\qquad &= \Ptker{a}{c} \Ptker{c}{b} {N \over \sqrt{k (N-k)} }
  \bigg( {k\over N-k} \bigg)^{\textstyle {n-1\over 2} }
    \exp\!\!\bigg[ - {k\over N-k} { |\vc{c}-\vc{b}|^2 \over 2 t} \bigg] \\
&\qquad \times
  \bigg( {N\over k} \bigg)^{\mbox{$n$$-$$1$}}
   \exp\!\!\bigg[-{N-k\over k} {|\vc{a}-\vc{c}|^2 \over 2 t}\bigg]. 
\end{split} \label{e:2ndeq}
\end{align}
Assume now that $n>1$ and
consider first the case $1\le k\le {N\over 2}$.  Since,
$t^s e^{-t r} \le \left( {s \over r e} \right)^s$
for all positive $r$, $s$ and $t$, the expression (\ref{e:1steq})
has an upper bound
\[
\Ptker{a}{c} \Ptker{c}{b} {N \over \sqrt{k (N-k)} } 2^{n-1} 
  \left( {(n-1)t\over e |\vc{a}-\vc{c}|^2} 
         \right)^{\textstyle {n-1\over 2} }.
\]
On the other hand, if ${N\over 2}\le k \le N-1$, then
the same reasoning can be applied to (\ref{e:2ndeq}) leading to the
same bound apart from the change of $|\vc{a}-\vc{c}|$ to 
$|\vc{c}-\vc{b}|$.
Since in both cases
these upper bounds are positive, the terms under inspection 
are for all $1\le k\le N-1$ bounded by
\[
\Ptker{a}{c} \Ptker{c}{b}
  \bigg( {4 (n-1) \over e} \bigg)^{\textstyle {n-1\over 2} } t^{n-1\over 2}
\left( {1\over |\vc{a}-\vc{c}|^{n-1}} + {1\over |\vc{c}-\vc{b}|^{n-1}} \right)
  {N \over \sqrt{k (N-k)} }.
\]

Now ${1\over N} \sum_{k=1}^{N-1} {\big[ {k\over N} 
  (1-{k\over N}) \big]^{-{1\over 2}} }$ 
is a lower Riemann sum for the integral
$\int_0^1 ds  {1\over\sqrt{s(1-s)}}=\pi$ and, therefore for all $N$,
\begin{equation}\label{e:sumbound}
{1\over N} \sum_{k=1}^{N-1} {N \over \sqrt{k (N-k)} } \le \pi
\end{equation}
and the Lemma follows if we choose $C_n = \pi 2^{n-1} (n-1)^{n-1\over 2}
e^{-{n-1\over 2}}$.

The case $n=1$ is even simpler to prove, as then both of the
exponentials in (\ref{e:1steq}) can be majorized by one and the Lemma,
with the choice $C_1 = \pi$, follows from (\ref{e:sumbound}).\hfill $\Box$

{\em Proofs of Corollaries \ref{cor:intbounds} and \ref{cor:supbound}:}\/
Define the following two functions on $(0,\infty)$
\begin{gather*}
I_s(t) := \int\!\! D\vc{a}\, |\vc{a}|^s e^{-t V(\vc{a})},
\qquad J_s(t) := \int\!\! D\vc{a}\, |\vc{a}|^s \Ptker{0}{a},
\end{gather*}
where $s$ is a real parameter.  $J_s$ is finite for $s>-n$ and, in fact,
it is then continuous, since a change of variables and relying on 
$n$-dimensional spherical coordinates shows that
\begin{equation}\label{e:js}
J_s(t) = ( 2 t )^{s/2} 
  {\Gamma\!\left({n+s\over 2}\right) \over \Gamma\!\left({n\over 2}\right)}.
\end{equation}
Under the assumptions of the theorem, also $I_0$ and $I_r$ are everywhere
finite.
They are then also continuous, since for every converging sequence 
$(t_k)$ in $(0,\infty)$ dominated convergence can be used to prove
$\lim_{k\to\infty} I_s(t_k) = I_s(\lim t_k)$.

The rest is just a matter of bookkeeping when integrating both sides of
the inequality (\ref{e:knl}).  Using 
Jensen's inequality for $r\ge 1$ and the obvious relation 
$(1+|\vc{a}|)^r - |\vc{a}|^r \le 1$ for $0\le r < 1$
it is easy to prove that for $r\ge 0$ and $\vc{a}$, $\vc{b}\in\Rn$ 
\[
|\vc{a}|^r \le 2^r ( |\vc{a}-\vc{b}|^r+|\vc{b}|^r).
\]
Applying this result, equation (\ref{e:js}) and 
a little algebra shows that
\begin{gather*}
\int\!\! D\vc{a}\, |\vc{a}|^r \knker{a}{a} \le
  C^{(1)} t^{-{n\over 2}} I_r(t)
  + C^{(2)} t^{{r-n\over 2}} I_0(t), \\
\int\!\! D\vc{a}D\vc{b}\, |\vc{a}|^r \knker{a}{b} \le
  C^{(3)} I_r(t) + C^{(4)} t^{r\over 2} I_0(t),
\end{gather*}
where the constants $C^{(k)}$ depend only on $n$ and $r$.  The functions
on the right hand sides are continuous and can therefore be chosen as
$F_r$ and $G_r$.

Since $e^{-tV(\vc{c})} \Ptker{c}{b}\le (2\pi t)^{-n/2}$,
Corollary \ref{cor:supbound} is an immediate consequence of
(\ref{e:js}) and Lemma \ref{lem:knl}. \hfill $\Box$

\begin{lemma}\label{lem:trh}
If $\ope{A}$ is bounded, $\ope{T}$ is Hilbert-Schmidt and $N\ge 1$, then
\[
\left|\tr\!\!\left(\ope{A} \T^{2 N}\right)\right| \le \norm{\ope{A}}
\tr(\T^*\T)^N.
\]
\end{lemma}
\begin{proof}
By the Schwarz inequality,
\[
\left|\tr\!\!\left(\ope{A} \T^{2 N}\right)\right|^2 \le
\tr\!\!\left[ (\ope{A}\T^N)^* (\ope{A}\T^N)\right]
\tr\!\!\left[ (\T^N)^*\T^N\right]
\]
and since $\tr\!\!\left[ (\ope{A}\T^N)^* (\ope{A}\T^N)\right]
 = \tr\!\!\left[ |\ope{A}|^2 \T^N (\T^N)^*\right]\le
 \norm{\ope{A}}^2 \tr\!\!\left[ (\T^N)^* \T^N \right]$, we get
\[
\left|\tr\!\!\left(\ope{A} \T^{2 N}\right)\right| \le
\norm{\ope{A}} \tr\!\!\left[ (\T^N)^*\T^N\right].
\]

Using a method similar to the so called min-max principle,
K.~Fan has shown\cite{fan49}
that for any finite-dimensional operator
(i.e.\ matrix) $\ope{B}$, $\tr\!\!\left[ (\ope{B}^N)^*\ope{B}^N\right]\le
\tr (\ope{B}^*\ope{B})^N$.  The finite-dimensional operators are dense in 
${\cal T}_2$ = the Hilbert space of Hilbert-Schmidt operators with the
inner product $(\cdot,\cdot)_{\rm HS}$.  Since the product, the inner
product and the norm are continuous on ${\cal T}_2$ and
$\tr\!\!\left[ (\ope{B}^N)^*\ope{B}^N\right] = \norm{\ope{B}^N}_{\rm HS}^2$, 
$\tr (\ope{B}^*\ope{B})^N = 
  (\ope{B}^*\ope{B}, (\ope{B}^*\ope{B})^{N-1})_{\rm HS}$, 
it is obvious that the result of Fan is valid for every Hilbert-Schmidt
operator. 
\end{proof}

\section{Discussion}
The main results of this paper are the two theorems,
\ref{th:realt} and \ref{th:complexb}.  The first one justifies the faith
of the lattice community that also expectation values of polynomials are
measurable with the lattice Monte Carlo methods if the potential
increases sufficiently fast at infinity.  The second one establishes the
possibility of using the lattice integrals for computations of the
expectation values even when the temperature has an imaginary
part.

There are a few shortcomings of the present results.  First, the exact
class of potentials for which the integral kernels $K_N$ can be used was
not determined.  This is not a serious problem since, by accepting the
necessity for a double limit, statistics given by any
potential in $L^1_{\text{loc}}$ can be
approached with the lattice integrals by replacing $V$ by a sequence
$V_k\in C^\infty$ which converges to it pointwise almost
everywhere.  The second shortcoming is the class of observables in 
Theorem \ref{th:complexb}:  the present proof cannot be easily extended
to include also polynomial observables for polynomial potentials,
although it is very natural to expect the integrals to converge to the
right limit by Theorem \ref{th:realt}.  Finally, the necessity of
using only even number of lattice points has no natural explanation and
as such it is expected to be an artifact of the present approach. 
Indeed, as has been noted earlier, the
requirement of even lattice sizes can be dropped
for continuous potentials.

\acknowledgements

I would like to thank A.~Kupiainen for enlightening
discussions and comments and for
suggesting several useful references.  I am also grateful to
M.~Chaichian and A.~Demichev for their comments.

\end{document}